\documentclass[12pt,preprint]{aastex}

\slugcomment{accepter on 1 March, 2003, to Astrophys. J.}

\shorttitle{ERO R1, an S0-like galaxy at $z\sim$1.5}
\shortauthors{Iye et al.}

\begin{document}

\title{ERO R1 in the field of Cl~0939+4713 \\
 - Evidence for an S0-like galaxy at $z\sim$1.5\altaffilmark{1}}

\author{Masanori \textsc{Iye}\altaffilmark{2,3}, Kazuhiro {\sc 
Shimasaku}\altaffilmark{4}, Satoshi {\sc Miyazaki}\altaffilmark{5}, \\ 
Chris {\sc Simpson}\altaffilmark{5}, Masatoshi {\sc Imanishi}\altaffilmark{2}, 
Nobunari {\sc Kashikawa}\altaffilmark{2,3}, \\ 
Tadayuki \textsc{Kodama}\altaffilmark{6}, 
Masashi {\sc Chiba}\altaffilmark{2,3}, Yoshihiko {\sc Saito}\altaffilmark{2}, \\
Miwa {\sc Goto}\altaffilmark{7},  
Fumihide {\sc Iwamuro}\altaffilmark{8}, Naoto {\sc Kobayashi}\altaffilmark{5}, \\
Sadanori \textsc{Okamura}\altaffilmark{4}, 
 and Hiroshi {\sc Terada}\altaffilmark{5}}
\altaffiltext{2}{\it Optical and Infrared Astronomy Division, National 
Astronomical Observatory, Mitaka, Tokyo 181-8588}
\email{\it iye@optik.mtk.nao.ac.jp}
\altaffiltext{3}{\it Department of Astronomy, School of Science, Graduate 
University for Advanced Studies, Mitaka, Tokyo 181-8588}
\altaffiltext{4}{\it Department of Astronomy, School of Science, University of 
Tokyo, Bunkyo-ku, Tokyo 113-0033}
\altaffiltext{5}{\it Subaru Telescope, National Astronomical Observatory, 650 
North A`Ohoku Place, Hilo, HI 96720, USA}
\altaffiltext{6}{\it Theoretical Astronomy Division, National Astronomical 
Observatory, Mitaka, Tokyo 181-8588}
\altaffiltext{7}{\it Institute for Astronomy, University of Hawaii, North 
A`Ohoku Place, Hilo, HI 96720, USA}
\altaffiltext{8}{\it Department of Astronomy, Kyoto University, Kita-shirakawa, 
Kyoto 606-8502}
\altaffiltext{1}{Based on data collected at Subaru Telescope, which is
operated by the National Astronomical Observatory of Japan, and in part on
observations with the NASA/ESA \textit{Hubble Space Telescope\/}, obtained
from the data archive at the Space Telescope Science Institute, which is
operated by AURA, Inc.\ under NASA contract NAS5-26555.} 

\begin{abstract}
We present further observations of the extremely red object
ERO~J094258+4659.2, identified by \citet{iye00} as ERO R1 in their deep images
of the cluster A851. We estimate its redshift independently by eight-band
photometric redshift determination and cross-correlation of a new $H$-band
spectrum with the optical spectra of local E/S0 galaxies, and conclude that
it lies at $z \sim 1.5$. Although its colors are consistent both with an
elliptical galaxy and an S0 galaxy at that redshift, its elongated shape
and exponential luminosity profile suggest the presence of an evolved
stellar disk component. We rule out the possibility that these properties
are strongly influenced by gravitational lensing by the foreground cluster,
and therefore conclude that this object is more likely to be an S0-like
galaxy, rather than a lensed elliptical.  The $H$-band spectrum does not
show strong H$\alpha$ emission and the star formation rate therefore
appears to be very modest.  The presence of such a galaxy with an
apparently relaxed disk of stars at this high redshift provides a new and
strong constraint on theoretical models which aim to explain the formation
and evolution of galaxies.
\end{abstract}

\keywords{galaxies: clusters: individual(A851)---galaxies: ellipticals and 
lenticulars, cD---galaxies: evolution}

\section{Introduction}

The cluster of galaxies A851 (= Cl~0939+4713) at $z=0.4$ was observed
during the first light period of Subaru Telescope using the optical imager,
Suprime-Cam, and the infrared imager, CISCO, both mounted at the Cassegrain
focus \citep{iye00}. Five EROs with $R-K' > 5$ were found in the field of
the cluster, and the object identified as ERO R1 was noted for its
outstanding color ($R-K'=7.4$) and elongated shape ($b/a = 0.4$) with its
major axis aligned with the equipotential surface of the cluster.  This
object, formally named to as ERO~J094258+46592 (but referred to as ERO~R1
throughout this paper for brevity), was first noted in the literature by
\citet{sma99} as object \#333 but did not receive special mention. In
Figure~\ref{fig:images} we present postage stamp images of ERO R1 in various
bands.

Preliminary analysis by \citet{iye00} from its \textit{RJK$'$} magnitudes
suggested that ERO~R1 is likely to be a passively-evolving early-type
galaxy at $1.0<z<1.6$ with modest dust extinction ($E(B-V) < 0.5$).  The
direction of the galaxy's major axis suggested that it might be the
gravitationally lensed image of a rounder background galaxy, although
the smooth and symmetrical shape argued in favor of intrinsic elongation.
\citet{iye00} therefore suggested that ERO~R1 might be an S0 or disk galaxy
at high redshift.

The existence of early-type galaxies with dynamically-relaxed stellar
disks, as ERO~R1 appears to be, is an important issue for galaxy formation
models to deal with. Since few EROs have been studied in detail, we have
made further observations of this object in an attempt to determine its
properties unambiguously and determine its relevance and importance in our
understanding of the ERO phenomenon and galaxy formation. We report on the
results of these observations in this paper. Unless otherwise stated, we
adopt $H_0 = 70\rm\,km\,s^{-1}\,Mpc^{-1}$, $\Omega_{\rm m}=0.3$, and
$\Omega_\Lambda=0.7$.

\section{Observations and reduction}

\subsection{IRCS imaging}

The InfraRed Camera and Spectrograph (IRCS; \citealt{ircs}) was used to
observe ERO~R1 on the night of UT 2001 Feb 3. A total of 30\,minutes
on-source integration was spent at $H$-band, and 40\,minutes at
\textit{zJ\/}-band (the \textit{zJ\/}-band filter is actually designed as an 
order-sorting
filter with a central wavelength of 1.11\,\micron\ and a bandwidth of
0.15\,\micron). These observations were made in groups of five 2-minute
integrations at different dither positions. The sky was clear and the
seeing was 0\farcs45--0\farcs60. The standard F9 star FS~127 \citep{haw01} was
observed as a photometric calibrator, with its magnitude estimated to be
$zJ = 12.05$ from its measured J(FS127)=11.95 and the color correction
of $zJ(FS127)-J(FS127)= 0.10$ assuming the typical color of $z-J=0.152$
for an F9 star and the effective central wavelengths of 0.80, 1.11. and 1.25
$\mu$m for the $z-$, $zJ-$, and $J-$band, respectively.  The estimated 
error of this zero-point calibration could be as large as 0.10, considering
the uncertainty in atmospheric extinction and the slight color deviation of
FS127 in $J-H$ and $J-K'$ with those for typical F9 stars.

The images were reduced with IRAF\footnote{IRAF is distributed by the
National Optical Astronomy Observatories, which are operated by the
Association of Universities for Research in Astronomy, Inc (AURA), under a
cooperative agreement with the National Science Foundation.} using standard
reduction procedures. First, bad pixels were identified and their values
replaced with values interpolated from the surrounding pixels. The frames
were then dark-subtracted and scaled to have the same median pixel value;
these scaled frames were then median-filtered to produce a flatfield
image. The individual dark-subtracted frames were then divided by a
normalized version of this flatfield, aligned by using a bright source near
ERO~R1, and combined to produce the final image.

\subsection{IRCS spectroscopy}

An $H$-band grism spectrum of ERO~R1 was taken with IRCS on the same night.
A 0\farcs9-wide, 20\arcsec-long slit, providing a spectral resolution $R
\sim 100$, was aligned with the target's major axis at ${\rm PA}=139\deg$.
The target was nodded between two positions on the slit separated by 7\arcsec;
each exposure was 5\,minutes, and 22 such pairs of exposures were taken to
provide a total integration time of 220\,minutes. The stars HR~1729,
HR~3928, and HR~4761 were also observed as spectrophotometric standards in
the same manner as the target.

Reduction was again performed in a standard manner within IRAF. After
subtracting the exposures in each pair from each other, the signal was
extracted in a 2\arcsec\ aperture. Wavelength calibration was performed
with a combination of night sky lines and an argon arc lamp spectrum.
Atmospheric absorption features were removed by dividing the spectrum of
the target by that of HR~4761, which was modeled as a $T=6200$\,K
blackbody. Finally, the flux scale was tied to the $H$-band photometry.

\subsection{Aperture Photometry}

Additional multi-color optical imaging of the A851 field was made by
\citet{kod01} in the \textit{BVRI\/} bands, using Suprime-Cam at the prime
focus of Subaru Telescope. For ease of reference, we list in
Table~\ref{tab:obslog} all the Subaru imaging data used in this paper. We
also retrieved the \textit{Hubble Space Telescope\/} NICMOS F160W-band
(approximately $H$) image of \citet{sma99} from the \textit{HST\/} data
archive. The photometric zeropoint of this image was tied to the published
magnitude of an unblended galaxy in the image.

Photometry of ERO~R1 was performed within a 4\arcsec-diameter circular
aperture in all the available images (including those presented in
\citealt{iye00}), using identical parameter settings of SExtractor
\citep{sex}. The magnitudes of ERO~R1 quoted in \citep{iye00} were obtained
in a different aperture and are therefore slightly, but not dramatically,
different from the values quoted here. We present the results of our
photometry in Table~\ref{tab:photom}. The poor seeing of the prime-focus
$B$-band image precluded accurate photometry of ERO~R1 in this band. No
aperture corrections were applied since the aperture is significantly
larger than the seeing in all cases, and therefore the measured colors
should be accurate\footnote{In fact, our aperture contains $\sim 98$\% of the total
galaxy flux, based on the exponential disk model we fit to the luminosity
profile in the next section.}.

\section{Properties of ERO R1}

\subsection{Spectral energy distribution}

The Hyperz code \citep{rod00,bol00} was used to estimate the redshift of
ERO~R1 from our multi-band photometry. Redshifts $z \leq 3$ and extinctions
$E(B-V) \leq 3.0$ (adopting the extinction curve of \citealt{cal00}) were
considered, and template SEDs corresponding to E, S0, Sa, Sc, and Im
galaxies, as well as a single starburst model, were tested. The
best-fitting SED (nominal probability of 57\%) was a 1-Gyr-old E/S0 galaxy
at $z=1.46$ with $A_V=1.4$\,mag, which clearly provides an excellent fit to
the data. The estimated formal error on the redshift is as small as 
0.03 (0.11) at 90\% (99\%) confidence. However, a second solution 
(nominal probability of 26\%) was found of a 1,4-Gyr-old E/S0 
with $A_V=0.8$\,mag at $z=1.59$, and the weighted mean
redshift is $z=1.49$.  We, therefore, assign the best photometric redshift
estimate at $z=1.5 \pm0.1$.

The extinction we derive, $A_V \lesssim 1.4$\,mag, is lower than that found
for dusty starburst galaxies, one of the two major populations of EROs.  On
the other hand, we fail to see any rest-frame UV upturn in our $V$-band
photometry, as is seen in local elliptical galaxies. However, the UV upturn
in nearby ellipticals arises mainly from evolved AGB stars which would
not be a significant population in a galaxy at $z \sim 1.5$. The location
of ERO~R1 in the $(R-K)$ vs $(J-K)$ two-color diagram also suggests that
this object is a passively evolving early-type galaxy, rather than a dusty
starburst \citep{poz00,man01}. This is further supported by the absence of
a redshifted H$\alpha$ line in our $H$-band spectrum, as we report in the
next section.

\subsection{$H$-band spectrum}

The pixel scale of IRCS (0\farcs06\,pixel$^{-1}$) is designed for high
spatial resolution imaging and spectroscopy, and is therefore not optimal
for observations of extended objects such as ERO~R1 (although it was the
only instrument available at the time of our observations). As a result,
the signal-to-noise ratio of our final spectrum is rather poor. We
present a binned version of this spectrum in Figure~\ref{fig:hspec}.

No strong emission lines are seen in the spectrum, although there is a
modest ($1.7\sigma$) enhancement at 1.72\,\micron, which could be
attributed to H$\alpha$ emission at $z=1.62$. In any case, we can place an
upper limit to the H$\alpha$ emission at any redshift $1.3 < z < 1.7$ of
$2.1 \times 10^{-16}\rm\,erg\,s^{-1}\,cm^{-2}$ ($3\sigma$), which
corresponds to a star formation rate of
$30h_{70}^{-2}\rm\,M_\odot\,yr^{-1}$ \citep{cim99, soi99, gla00}.
Considering the modest amount of extinction derived in the previous
section, we conclude that any starburst activity in this galaxy must be
very modest.

We attempted to determine the redshift of ERO~R1 independently by performing
a cross-correlation analysis of our $H$-band spectrum with redshifted model
galaxy spectra. The average level of SED was subtracted from 
the measured $H$-band SED and it was then cross correlated with a similarly
bias-subtracted model SED of a redshifted early-type galaxy chosen from a
model library \citep{kod98}.  The resulting cross correlation moment was calculated
for a redshift range $1.0 < z <2.5$ with an interval of 0.005 for models with various
age.   The root-mean-square of the derived moment is taken as one 
sigma error for the assessment of significance.
Some example results of this analysis for S0 galaxies are presented in
Figure~\ref{fig:xcorr}. The most significant peak ($3.3\sigma$) was found
for a 3-Gyr-old single-burst model at $z=1.50\pm0.02$, with a second peak
($2.8\sigma$) at $z=1.72\pm0.02$. Models for ellipticals give similar results
but with somewhat less significance.
Note that a 1-Gyr-old E/S0 model does not
produce strong peaks in this correlation function.  This is apparently due to 
the fact that the TiO absorption features have not had time to develop for
early-type galaxies. This analysis therefore suggests
, though naively, that ERO~R1 is at least $\sim 2$\,Gyr old.

Although the template does not look like a good match to the observed SED
of low signal-to-noise ratio, we show the best fit S0 SED model, 
a 3-Gyr-old single-burst model at $z=1.50$, overlaid on the observed 
SED in Figure~\ref{fig:hspec}, to illustrate the TiO absorption features 
which lie in the 600--800\,nm (rest-frame) region.

\subsection{Luminosity profile}

We use our $K'$-band image of ERO~R1 to study its structure, since this has
the highest signal-to-noise ratio, was taken in the best seeing, 
samples rest-frame wavelengths which are less affected by dust extinction
and any transient star formation, and reflects the distribution of stellar population 
most appropriately among the observed bands.

At the isophotal limit of $\mu_{K'} = 22\rm\,mag\,arcsec^{-2}$, the galaxy
has a size of $3\farcs7 \times 1\farcs8$ ($31 \times 15 h_{70}^{-1}$\,kpc
at $z=1.5$), which is fairly large. We performed isophotal fitting to
determine the luminosity profile, and Figure~\ref{fig:lumprof} shows that
this is consistent with an exponential profile convolved with 0\farcs45
seeing. We determined that the effects of seeing on the observed profile
are negligible for $r > 0\farcs5$, and therefore our result is not
dependent on the accuracy with which we had measured the seeing in our
image. Similarly, reasonable uncertainties in the background subtraction do
not affect the profile at isophotal levels brighter than
22.5--$23\rm\,mag\,arcsec^{-2}$. The scale length of the disk measured in
the $K'$-band ($\lambda_{\rm rest} = 0.85\,\micron$) is $\alpha^{-1} =
3h_{70}^{-1}$\,kpc, which is consistent with the distribution of disk scale
lengths measured by \citet{lil98} for galaxies from the CFRS/LDSS
surveys. They found that 48/104 galaxies with $0.2 < z < 1.0$ had scale
lengths (in the F814W filter) $\alpha^{-1} > 2.2h_{70}^{-1}$\,kpc. If
ERO~R1 evolves passively to the present epoch, its absolute magnitude would
be $M_K = -25.0 + 5 \log h_{70}$, or about $3L^*$ \citep{gar97}.
Figure~\ref{fig:kormendy} shows the absolute total $K$ magnitude $M_K$
plotted against the effective radius $R_{\rm e}$ of E/S0 galaxies in the
Coma cluster \citep{pah99}, with ERO~R1 added. The size and the absolute
total magnitude of R1 appear to be similar to those of giant E/S0 galaxies.

In an effort to further classify the morphological type of ERO~R1, we
attempted to perform bulge--disk decomposition of the object's luminosity
profile. The central excess above the seeing-convolved exponential disk
profile might argue for a modest bulge component. A detailed analysis is
not possible given the relatively small size of the object and our
seeing-limited images, so instead we attempted to fit the profile with five
models, where the bulge component contributes 0\%, 30\%, 50\%, 70\%, and 100\% of
the total $K'$-band light. We found that the model with $L_{\rm
bulge}/L_{\rm total} = 0.3$ provides the best fit to the data over the entire range.
The $L_{\rm bulge}/L_{\rm total} = 1.0 and 0.7$ models, and probably the 0.5 
model, are brighter than the observed profile both at the central region and the 
outer region, while the pure disk model with $L_{\rm bulge}/L_{\rm total} = 0.0$ 
gives surface brightness fainter than was observed at the central region. 
We therefore conclude that ERO~R1
is similar to nearby S0 galaxies with small bulges. Note that gravitational
lensing may deform the shape of the image but will not change the
 de Vaucouleurs profile into an exponential profile.
Similar analysis on HST $F160W$ image and CISCO $H$-band image gave consistent
result with that from the above $K'$-band analysis.

 \citet{mor00} made a morphological study of 41 EROs. Most of these objects
appeared to be ellipticals, but a few galaxies with compact morphologies
were best fitted with an exponential light profile. ERO~R1 has an elongated
shape comparable to the most elongated object in their sample (ID 22), but
R1 appears to possess a smoother light profile. It is not clear how much of
this may be attributable to the higher resolution and somewhat poorer
signal-to-noise ratio of their \textit{HST\/} data.

\section{Discussion}

\subsection{Clustering at $z=1.5$?}

\citet{dad00} found a strong clustering signal in their sample of 400 EROs
and argued as a result that the ERO population is dominated by
high-redshift elliptical galaxies, rather than dusty starbursts. Since
ERO~R1 appears to be a member of the class of passively-evolving early-type
galaxies, it may belong to some element of large-scale structure and
therefore there may be an overabundance of galaxies at $z \sim 1.5$. The
Suprime-Cam imaging of Cl~0939+4713 by \citet{kod01} was intended to study
the large-scale structure in the distribution of galaxies, and a
preliminary (unpublished) photometric redshift analysis based on their
\textit{BVRI\/} data shows a slight excess of galaxies at $z=1.5$. However
the excess is very modest and the uncertainty in photometric redshifts
determined solely from optical data is rather large at $z>1$. Further
studies, including wider-field infrared imaging, are clearly needed before
a firm statement about the presence or absence of a distant cluster can be
made.


\subsection{Evaluation of possible lensing effects}

ERO~R1's major axis is almost perpendicular to the direction toward the
surface mass density peak of the cluster as derived from the weak lensing
analysis of \citet{iye00}. This led \citet{iye00} to suggest that the
elongated nature of ERO~R1 could be due to image shear of an intrinsically
rounder galaxy, caused by the strong gravitational field of the intervening
cluster. \citet{sei96} performed a weak-lensing analysis of Cl~0939+4713
and found that the cluster mass distribution closely traces the
distribution of bright member galaxies. High-resolution X-ray observations
by \citet{sch98} confirmed the presence of concentrations of hot plasma
consistent with the distribution of mass and galaxies, and indicate that
this cluster is still actively evolving towards a virialized state.

The proximity of ERO~R1 to the cluster center, whether defined as the
optical center ($\alpha=09^{\mathrm h}42^{\mathrm m}56\fs9$,
$\delta=+46^{\circ}59\arcmin23\farcs1$, 21\arcsec\ from R1) or the X-ray peak
M1 ($\alpha=09^{\mathrm h}42^{\mathrm m}58\fs2$,
$\delta=+46^{\circ}58\arcmin52\arcsec$, 20\farcs4 from R1) makes it conceivable
that the object's morphology suffers considerable gravitational distortion.
In order for the light from a galaxy at a projected distance $r$ from the
cluster center to be strongly lensed, a lensing mass with radius $r$ of
$M(<r) > \pi \sigma_{\mathrm cr} r^2$ is required, where the critical
surface mass density, $\sigma_{\rm cr}$, depends on the distances to the
lens and the background source, as well as the cosmological model. Since we
know the redshifts to the source ($z=1.5$) and lens ($z=0.4$), we can
calculate the required lensing mass, which is $M(<r) = 1.4 \times 10^{16}
r_{\mathrm Mpc}^2 M_\odot$, or $3 \times 10^{13} M_{\odot}$ for $r =
44$\,kpc, assuming the center of the cluster to be the location where the
reconstructed surface mass density peaks (see Figure~13 of
\citealt{iye00}).

Figure~\ref{fig:massprof} shows the integrated radial mass profile of the
cluster derived from the weak lensing analysis using the method described
in \citet{kai95}. Adopting the surface mass density peak as the center of
the cluster, we have performed the calculation for two values of the mean
redshift of the background galaxies, $z_{\mathrm s}=1.0$ and $z_{\mathrm
s}=3.0$. Although we cannot evaluate exactly the value of $z_{\mathrm s}$,
it is likely to be closer to 1.0 than 3.0.
The integrated mass of the cluster CL~0939+4713 was estimated using X-ray 
observation by Schindler et al.(1998) and they deduced a mass of 
$\sim 1.5 \times 10^{13} M_\odot$ at 100 kpc from the X-ray peak M1 
whose location coincides with the mass peak deteced by lensing.  
The present mass estimate from lensing, as is shown in Figure 7, 
is consistent with their result.

From this figure, the mass wihin the projected
distance to ERO~R1 is at most $(1.0 \pm 0.4) \times 10^{13} M_{\odot}$, or
a factor of 3 lower than the mass required for strong lensing. We made
additional estimates of the mass after displacing the assumed cluster
center in four directions, but found that the result was changed by less
than 30\%. Note in particular that if the cluster center is at the X-ray
peak M1, the distance to R1 increases and the required mass
becomes even larger. We therefore conclude that the cluster mass is
insufficient to strongly lens the light from ERO~R1.

Since the distance of ERO R1 from the cluster center is relatively small, the image, 
if lensed, would
 show distinct bending, whose radius of curvature is
approximately equal to the distance of ERO R1 from the lens center.
The observed image of ERO R1 does not show any  expected  curvature and hence also 
is not consistent with  strong lensing interpretation.

Although ERO~R1 is not strongly lensed, the large mass of the intervening
cluster may affect its observed shape, and we have attempted to reconstruct
its unlensed image. We model the cluster as an isothermal sphere, using
Schindler et al.'s (1998) estimate of the total mass, for five different
assumed locations of the center of mass. Figure~\ref{fig:unlensed} (Case~0)
presents the unlensed image if the center of mass is located at the X-ray
peak M1. In Figure~\ref{fig:unlensed4}, we present four additional images
where we have displaced the center of mass by 10\arcsec\ (equal to the
smoothing applied to the X-ray image) from M1; we refer to these as Cases
N, W, E, and S, according to whether the displacement was made to the
north, west, east, or south, respectively. The calculation for Case~N
requires data in an unobserved region, and we assigned the lowest observed
surface brightness to this region. We constructed the luminosity profiles
of the five unlensed images and confirmed that they were all well-fit with
an exponential disk profile, as expected. ERO~R1 therefore possesses a
significant disk component and its elongated shape is not due to
gravitational lensing of a rounder elliptical galaxy.

\subsection{Implications for the early formation of an S0 galaxy}

Our interpretation of an apparently relaxed, massive S0 galaxy at $z=1.5$
is made difficult since we do not know whether it lies in a cluster or in
the field. The luminous early-type populations of rich clusters are known
to be remarkably homogeneous and the bulk of their stellar populations
appear to be old \citep{bow92, ell97}. \citet{jon00} found from their
spectroscopic data that there is no statistically significant difference
between the luminosity-weighted age of the E and S0 galaxies in clusters at
$0.37 < z < 0.56$, supporting the conventional formation model of old,
coeval E and S0 galaxies. In nearby clusters, the density--morphology
relation and the formation of S0 galaxies are often related to the effects
of ram pressure stripping \citep{dre80}. This is unlikely to be the cause
of ERO~R1's morphology, however, since it is not clear whether ram pressure
stripping could occur effectively at such an early cosmic epoch, and our
optical--infrared images provide no evidence for the existence of a rich
cluster. We note that \citet{soi99} identified another ERO in the field of
Cl~0939+4713 (named Cl~0939+4713~B) as a passively-evolving galaxy at
$z=1.58$ (the redshifts of ERO~R1 and Cl~0939+4713~B are marginally
consistent given the uncertainties), although this is $\gtrsim 700\,$kpc
from R1 and therefore cannot be said to belong to the same `cluster'.

It is far more likely, therefore, that ERO~R1 is simply a field galaxy. A
morphological analysis of galaxies in the Hubble Deep Fields by
\citet{rod01} led these authors to conclude that massive E/S0 galaxies tend
to disappear from flux-limited samples at $z>1.4$. They therefore suggested
that the era $1<z<2$ corresponds to a very active phase for the assembly of
massive E/S0 galaxies in the field.

\citet{smi02} studied the gravitationally lensed ERO~J003707+0909.5 at
$z=1.6$ in the field of the cluster A68. From their reconstructed image and
photometric analysis, they concluded that this object is an $L^*$
early-type disk-galaxy, and therefore very similar to ERO~R1. In addition
to the known classes of EROs, i.e., dusty starbursts such as HR~10
\citep{hu94,gra96,dey99} and passively-evolving ellipticals,
\citeauthor{smi02} suggested that a significant fraction of EROs may evolve
into luminous spiral galaxies by the present epoch. 

If the absence of high over density of galaxies around R1 is confirmed, 
some constraints will be placed as to the physical mechanisms of forming
S0 galaxies.  As already discussed, ram-pressure stripping will not be a
dominant factor and instead a mechanism which can be effective in lower
density regions will be required.
This is consistent with the recent findings that the truncation of star
formation in galaxies is not the phenomenon restricted to the dense
cluster cores but is wide spread to relatively low density regions
(e.g., Kodama et al.2001; Balogh et al.1999; 2002, Lewis et al.2002;
 Gomez et al.2002).


\section{Summary}

We have presented new observations of ERO~R1 in the field of the rich
cluster CL~0939+4713 (A851). By independent analyses of its multi-color
photometry and $H$-band spectrum, we have concluded that it is a
passively-evolving galaxy at $z=1.5 \pm 0.1$. We have shown that
gravitational lensing by the foreground cluster may play some part in, but
cannot be the main cause of, its elongated shape. The radial luminosity
profile is well-represented by an exponential disk with a relatively small
bulge component, indicating that the stellar population in this galaxy is
well-relaxed. Our results have led us to conclude that ERO~R1 is an S0-like
galaxy whose stellar disk is already well-developed by $z=1.5$. This
conclusion poses a new constraint on models of disk formation at high
redshift.

\acknowledgments

We are grateful to the staff of Subaru Telescope for their support during
the observations reported in this paper.


\clearpage

\begin{figure}
\caption[]{$V$, $R$, $I$, $zJ$, $J$, $H$, $F160W$, and $K'$ images of ERO R1. The size 
of the images is $5.''8 \times 5.''8$. North is up and East is to the
left.}\label{fig:images}
\end{figure}
\clearpage

\begin{figure}
\caption[]{Broad-band photometry of ERO R1 (filled circles corresponding to
the $V$, $R$, F702W, $I$, \textit{zJ\/}, $J$, $H$, and $K'$ bands) compared
with the best-fit evolutionary synthesis model of a 1 Gyr E/S0
galaxy at $z=1.46$ with $A_V=1.4$ mag of extinction.}\label{fig:sed}
\end{figure}
\clearpage

\begin{figure}
\caption[]{$H$-band spectrum of ERO R1 (filled circles with error bars)
overlaid on a 3 Gyr single burst passive evolution model at redshift 1.5
(broken line). The ordinate is normalized by $F_{\lambda} = 2 \times
10^{-17}\,erg\,s^{-1}\,cm^{-2}\AA^{-1}$.}\label{fig:hspec}
\end{figure}
\clearpage

\begin{figure}
\caption[]{Cross-correlation of the $H$-band spectrum with single burst
passive evolution models of galaxies at various ages.
The range of the redshift in the last panel is terminated at $z=2.1$,
corresponding to the largest possible redshift for a 3-Gyr-old
galaxy.}\label{fig:xcorr}
\end{figure}
\clearpage

\begin{figure}
\caption[]{$K'$-band isophotal luminosity profile of ERO~R1 
(filled circles). The solid and dotted curves represent 
seeing-convolved profiles 
for a pure exponential disk and for a pure elliptical galaxy with the 
de Vaucouleurs profile,
respectively.  The broken curve shows the seeing convolved best fit 
composite profile with $L_{bulge}/L_{total}=0.3$.}\label{fig:lumprof}
\end{figure}
\clearpage

\begin{figure}
\caption[]{A plot of $K$-band absolute total magnitude $M_K$ against effective
radius $R_{\mathrm e}$ for galaxies in the Coma cluster (small circles).
The large square indicates the location of ERO~R1 in this plot, which is
consistent with the Coma galaxies.}\label{fig:kormendy}
\end{figure}
\clearpage

\begin{figure}
\caption[]{The radial mass profile of Cl~0939+4713 estimated from the
observed tangential shears of faint ($R>24$) galaxies for two assumed mean
redshifts, $z_{\mathrm s} = 1.0$ (filled circles) and $z_{\mathrm s} = 3.0$
(triangles). The error bars are obtained using orthogonal shear components
and do not reflect uncertainties in $\Sigma_{c}$. The \citet*{nfw96} mass
profile with $V_{200} = 1500\rm\,km\,s^{-1}$ and $c = 3.0$ is presented as
a dashed line for reference.}\label{fig:massprof}
\end{figure}
\clearpage

\begin{figure}
\caption[]{Unlensed image of ERO R1 constructed by modeling the lensing
cluster as an isothermal sphere with its center located at the X-ray
intensity peak M1 (Case 0)}\label{fig:unlensed}
\end{figure}
\clearpage

\begin{figure}
\caption[]{Unlensed images of ERO R1 for four other cases where the center of
the cluster has been displaced by 10\arcsec\ in the N, E, S, and W
directions.}\label{fig:unlensed4}
\end{figure}
\clearpage

\begin{table*}
\small
\caption{Observation log of our Subaru data.\label{tab:obslog}}
\vspace{6pt}
\begin{tabular*}{\columnwidth}{@{\hspace{\tabcolsep}
\extracolsep{\fill}}cccccc}
\hline\hline\\[-6pt]
Date  &   Instrument(Focus) & Band  &  Exposure  &  Seeing  &  5 $\sigma$ \cr
Jan.21, 2001 & Suprime-Cam(Prime) & $B$ &  900 s $\times$ 4 & 1\farcs1 & 27.0 \cr 
Jan.22, 2001 & Suprime-Cam(Prime) & $V$ &  540 s $\times$ 4 & 0\farcs7 & 26.3 \cr 
Jan.12, 1999 & Suprime-Cam(Cass) & $R$ &  900 s $\times$ 4 & 0\farcs45 & - \cr 
Jan.21, 2001 & Suprime-Cam(Prime) & $R$ &  990 s $\times$ 4 & 1\farcs1 & 26.2 \cr  
Jan 22, 2001 & Suprime-Cam(Prime) & $I$ &  315 s $\times$ 4 & 0\farcs7 & 24.6 \cr 
Feb.2, 2001 & IRCS(Cass) & $zJ$ &  120 s $\times$  20 & 0\farcs55 & - \cr
Jan.13, 1999 & CISCO(Cass) & $J$ &   60 s $\times$ 48 & 0\farcs3 & - \cr 
Feb.2, 2001 & IRCS(Cass) & $H$ &  120 s $\times$  15 & 0\farcs55 & - \cr 
Jan.11, 1999 & CISCO(Cass) & $K'$ &   20 s $\times$  144 & 0\farcs3 & - \cr 
\hline
\end{tabular*}
\end{table*}

\begin{table*}
\small
\caption{Photometry of ERO R1.\label{tab:photom}}
\vspace{6pt}
\begin{tabular*}{\columnwidth}{@{\hspace{\tabcolsep}
\extracolsep{\fill}}ccc}
\hline\hline\\[-6pt]
Band  &  Magnitude  &  Error  \cr
$B$ Suprime-Cam & - & - \cr
$V$ Suprime-Cam & 26.48 & 0.22 \cr
$F702W$ WFPC2  & 24.86 & 0.06 \cr
$R$ Suprime-Cam & 25.59 & 0.05 \cr
$I$ Suprime-Cam & 23.78 & 0.12 \cr
$zJ$ IRCS & 21.5 & 0.24 \cr
$J$ CISCO & 20.47 & 0.01 \cr
$H$ NICMOS & 19.24 & 0.01 \cr
$H$ IRCS & 19.40 & 0.09 \cr
$K'$ CISCO & 18.32 & 0.01 \cr
\hline
\end{tabular*}
\end{table*}


\begin{thebibliography}{}


\bibitem[Balogh(1999)]{bal99} Balogh, M.L., Morris, S.L., Yee,H.K.C.,
Carlberg,R.G., Ellingson,E., 1999, \apj, 527, 54.

\bibitem[Balogh(2002)]{bal02} Balogh,M.L., Bower,R.G., Smail,I., Ziegler,B.L.,
Davies,R.L., Gaztelu,A., Fritz,A., 2002, \mnras, 337, 256.


\bibitem[Bertin \& Arnouts(1996)]{sex} Bertin, E., \& Arnouts, S. 1996,
\aaps, 117, 393

\bibitem[Bolzenella et al.(2000)]{bol00} Bolzenella, M., Miralles, J.-M.,
\& Pello, R. 2000, \aap, 363, 476

\bibitem[Bower et al.(1992)]{bow92} Bower,R., Lucey,J., Ellis,R., \ 1992,
\mnras, 254, 601


\bibitem[Calzetti et al.(2000)]{cal00} Calzetti, D., Armus, L., Bohlin,
R. C., Kinney, A. L., Koornneef, J., \& Storchi-Bergmann, T. 2000, \apj,
533, 682

\bibitem[Cimatti(1999)]{cim99} Cimatti, A., et al. 1999, \aap, 352, L45.

\bibitem[Daddi et al.(2001)]{dad00} Daddi, E., Cimatti, A., Pozzetti, C., 
Hoekstra, H., Roettgering, H. J. A., Renzini, A., Zamorani, G., \&
Mannuri, F. \ 2000, \aap, 361, 535

\bibitem[Dey et al.(1999)]{dey99} Dey, A., Graham, J. R., Ivison, R. J.,
Smail, I., Wright, G. S., \& Liu, M. C. 1999, \apj, 519, 610

\bibitem[Dressler(1980)]{dre80} Dressler, A. 1980, \apj., 236, 351

\bibitem[Dressler et al.(1997)]{dre97} Dressler, A., et al. 1997, \apj, 
490, 577

\bibitem[Ellis et al.(1997)]{ell97} Ellis, R., Smail, I., Dressler, A.,
Couch, W., Oemler, A., Butcher, H., \& Sharples, R. 1997, \apj, 483, 582

\bibitem[Fernandez-Soto et al.(2002)]{fer02} Fernandez-Soto,A., Lanzetta,K.M., Chen,H.W., Levine,B., Yahata,N., 2002, \mnras, 330, 889

\bibitem[Gardner et al.(1997)]{gar97} Gardner, J. P., Sharples, R. M.,
Frenk, C. S., \& Currasco, B. E. 1997, \apj, 490, L99

\bibitem[Glassman \& Larkin (2000)]{gla00} Glassman, T. M., \&
Larkin, J. E. 2000, \apj, 539, 570

\bibitem[Gomez(2003)]{gom03} Gomez,P.L., Nichol,R.C., Miller,C.J., Balogh,M.L., Goto,T.,
Zabludoff,A.I., Romer,A.K., et al., 2003, \apj, 584, 210.

\bibitem[Graham \& Dey(1996)]{gra96} Graham, J. R., \&  Dey, A.
1996, \apj, 471, 720

\bibitem[Hawarden et al.(2001)]{haw01} Hawarden,T ., Leggett, S. K.,
Letawsky, M. B., Ballantyne, D. R., \& Casali, M. M. 2001, \mnras, 325, 563
\bibitem[Hu \& Ridgway(1994)]{hu94} Hu, E. M., \& Ridgway, S. E.
1994, \aj, 107, 1303

\bibitem[Iye et al.(2000)]{iye00} Iye, M., et al. 2000, \pasj, 52, 9

\bibitem[Jones et al.(2000)]{jon00} Jones, L., Smail, I., \& Couch, W., J.
2000, \apj, 528, 118

\bibitem[Kaiser et al.(1995)]{kai95} Kaiser, N., Squires, G., \&
Broadhurst, T. 1995, \apj, 449, 460

\bibitem[Kobayashi et al.(2000)]{ircs} Kobayashi, N., et al. 2000, in
Proc.\ SPIE 4008: Optical and IR Telescope Instrumentation and Detectors,
eds M. Iye \& A. F. Moorwood, 1056 

\bibitem[Kodama et al.(1998)]{kod98} Kodama, T., Arimoto,N., Barger,A.J., \& Arag'on-Salamanca,A., 2001, \aap, 334, 99

\bibitem[Kodama et al.(2001)]{kod01} Kodama, T., Smail, I., Nakata, F.,
Okamura, S., \& Bower, R.G. 2001, \apj,  562, L9

\bibitem[Lewis(2002)]{lew02} Lewis,I., Balogh,M., De Propris,R., Couch,W., Bower,R., Offer,A., Bland-Hawthorn,J. et al. 2002, \mnras, 334, 673.

\bibitem[Lilly et al.(1998)]{lil98} Lilly, S., et al. 1998, \apj, 500, 75

\bibitem[Mannucci et al.(2001)]{man01} Mannucci, F., Pozetti, L.,
Thompson, D., Oliva, E., Baffa, C., Comoretto, G., Gennari, S., \& Lisi, F.
2001, \mnras, 327, L57

\bibitem[Moriondo et al.(2000)]{mor00} Moriondo, G., Cimatti, A., \&
Daddi, E. 2000, \aap, 364, 26

\bibitem[Navarro et al.(1996)]{nfw96}
Navarro, J., Frenk, C., \& White, S. 1996 \apj, 462, 563

\bibitem[Pahre (1999)]{pah99} Pahre, M. A. 1999, \apjs, 124, 127


\bibitem[Pozzetti and Mannucci(2000)]{poz00} Pozzetti, L., \& Mannucci, F.
2000, \mnras, 317, L17

\bibitem[Rodighiero et al.(2001)]{rod01} Rodighiero, G., Franceschini, A., \& 
Fasano, G. 2001, \mnras, 324, 491

\bibitem[Rodighiero et al.(2000)]{rod00} Rodighiero, G., Granato, G. L., 
Franceschini, A., Fasano, G., \& Silva, L. 2000, \aap, 364, 517

\bibitem[Schindler et al.(1998)]{sch98} Schindler, S., Bellon,i P.,
Ikebe, Y., Hattori, M., Wambsganss, J., \& Tanaka, Y. 1998, \aap \  338, 843

\bibitem[Seitz et al.(1996)]{sei96} Seitz, C., Kneib, J.-P., Schneider, P. \&
Seitz, S. 1996, \aap, 314, 707

\bibitem[Smail et al.(1997)]{sma97} Smail, I., Dressler, A., Couch, W. J.,
Ellis, R., Oemler, A., Butcher, H., \& Sharples, R. M., 1997, \apjs, 110, 213

\bibitem[Smail et al.(1999)]{sma99} Smail, I., Morrison, G., Gray, M. E.,
Owen, F. N., Ivison, R. J., Kneib, J.-P., \& Ellis, R. S. 1999, \apj, 525, 609

\bibitem[Smail et al.(2002)]{sma02} Smail, I., Owen, F. N., Morrison, G. E., 
Keel, W. C., Ivison, R. J., \& Ledlow, M. J. 2002, astro-ph/0208434

\bibitem[Smith et al.(2002)]{smi02} Smith, G. P., Smail, I., Kneib, J.-P.,
Davis,C. J., Takamiya, M., Ebeling, H., \& Czoske, O. 2002, astro-ph/0203402

\bibitem[Soifer et al.(1999)]{soi99} Soifer, B. T., Matthews, K.,
Neugebauer, G., Armus, L., Cohen, J. G., \& Persson, S. E. 1999, \aj, 118, 206

\bibitem[Susa \& Umemura(2000a)]{sus00a} Susa, H., \& Umemura, M.
2000, \mnras, 316, L17

\bibitem[Susa \& Umemura(2000b)]{sus00b} Susa, H., \& Umemura, M.
2000, ApJ, 537, 578



\end{thebibliography}
\end{document}